\documentclass[final,3p,times,twocolumn]{elsarticle}

\usepackage{amssymb}

\journal{Nuclear Instruments and Methods A}

\begin{document}

\begin{frontmatter}



\title{A method to localize gamma-ray bursts using POLAR}


\author[DPNC,ISDC]{E.~Suarez-Garcia \corref{cor1}}
\ead{estela.suarez@unige.ch}
\author[DPNC,ISDC]{D.~Haas}
\author[PSI]{W.~Hajdas}
\author[LAPP]{G.~Lamanna}
\author[DPNC]{C.~Lechanoine-Leluc}
\author[IPJ]{R.~Marcinkowski}
\author[PSI]{A.~Mtchedlishvili}
\author[DPNC]{S.~Orsi}
\author[DPNC]{M.~Pohl}
\author[ISDC]{N.~Produit}
\author[DPNC]{D.~Rapin}
\author[IPJ]{D.~Rybka}
\author[LAPP]{J.-P.~Vialle}

\cortext[cor1]{Corresponding author. Tel.: +41 22 379 6923. Fax: +41 22 379 6992}
\address[DPNC]{DPNC, 24 Quai Ernest-Ansermet, Universit\'{e} de Gen\`{e}ve, 1205 Gen\`{e}ve, Switzerland.\fnref{DPNC}}
\address[ISDC]{INTEGRAL Science Data Centre, 16 Chemin d'\'{E}cogia, 1290 Versoix, Switzerland \fnref{ISDC}}
\address[LAPP]{Laboratoire d'Annecy-le-vieux de Physique des Particules, 9 Chemin de Bellevue, F-74941 Annecy-le-Vieux, France \fnref{LAPP}}
\address[PSI]{Paul Scherrer Institut, 5232 Villigen PSI, Switzerland \fnref{PSI}}
\address[IPJ]{The Andrzej Soltan Institute for Nuclear Studies, 69 Hoza str., 00-681 Warsaw, Poland \fnref{IPJ}}

\begin{abstract}
The hard X-ray polarimeter POLAR aims to measure the linear polarization of the 50 -- 500 keV photons arriving from the prompt emission of $\gamma$-ray bursts (GRBs). The position in the sky of the detected GRBs is needed to determine their level of polarization. We present here a method by which, despite of the polarimeter incapability of taking images, GRBs can be roughly localized using POLAR alone. For this purpose scalers are attached to the output of the 25 multi-anode photomultipliers (MAPMs) that collect the light from the POLAR scintillator target. Each scaler measures how many GRB photons produce at least one energy deposition above 50 keV in the corresponding MAPM. Simulations show that the relative outputs of the 25 scalers depend on the GRB position. A database of very strong GRBs simulated at 10201 positions has been produced. When a GRB is detected, its location is calculated searching the minimum of the $\chi^2$ obtained in the comparison between the measured scaler pattern and the database. This GRB localization technique brings enough accuracy so that the error transmitted to the 100\% modulation factor is kept below 10\% for GRBs with fluence ${\rm F_{tot}} \ge 10^{-5}$ erg cm$^{-2}$. The POLAR localization capability will be useful for those cases where no other instruments are simultaneously observing the same field of view. 
\end{abstract}


\end{frontmatter}


\section{Introduction}
\label{sec:intro}

Gamma-ray bursts (GRBs) are short flashes of $\gamma$-rays likely produced during the creation of a black-hole at cosmological distances. In a few seconds a huge amount of energy between $10^{51}$ and $10^{53}$ erg is released such that GRBs are the most violent explosions in the Universe. This prompt emission presents a great variability of lightcurves and its spectrum follows a broken power law function, often described using the Band model \cite{band}. The Band model fits the spectrum using four parameters: an amplitude ($A$), the low- and high-energy spectral indexes ($\alpha$ and $\beta$, respectively), and the peak energy (E$_{\rm peak}$) of the power density spectrum $\nu F_{\rm \nu}$, which represents the total energy flux per energy band.


After the prompt $\gamma$-ray emission has finished, a long lasting afterglow can be observed at various longer wavelengths. The current theoretical picture is that GRBs are produced when a massive star collapses at the end of its life or when two compact objects merge, giving birth in either case to a black hole. The collapsar scenario would be responsible for the long duration \cite{woosley} and the merger scenario for short duration GRBs \cite{eichler}. The separation line between short and long GRBs was traditionally taken at 2 seconds \cite{kouveliotou}, but recently suggested rather at 5 seconds \cite{donaghy}. For recent reviews on GRBs, see e.g. \cite{woosleybloom, zhang, zhang2}.

Several theoretical models have been proposed to explain which processes are responsible for the GRB emission. In the standard \textit{fireball} model \cite{piran} the explosion produces a jet in which expanding shells cause internal shocks when a fast shell overtakes a slower moving one. In these shocks relativistic electrons are accelerated, generating through synchrotron emission the X- and $\gamma$-ray photons that we see as the GRB prompt emission. The resulting photon polarization is typically low ($\approx$ 10\%) reflecting the hydrodynamic nature of the model. In the \textit{electromagnetic} model \cite{lyutikov} the energy is extracted from the central engine  by an electromagnetic field, transported in a plasma to large distances by strongly magnetized wind, and dissipated through fast magnetic reconnections that generate the GRB radiation. Photons emitted in this process can present polarization as high as 50\%. In the \textit{cannonball} model \cite{dar} the $\gamma$-ray emission is generated by inverse Compton scattering of soft photons that hit the previously ejected material. The polarization level depends on the opening angle and Lorentz factor of the jet, predicting therefore all possible values between 0 and 100\% polarization.

Precise polarimetry measurements of the GRB prompt emission can distinguish between different theoretical models, offering unique information on the emission mechanism of the GRBs and on the composition and geometric structure of their jets \cite{lazzati, toma}. To date, only a few measurements of the prompt GRB polarization have been performed, all of them with instruments that had not been designed for this purpose and lacked in many cases a good characterization of their systematic effects, enough effective area, or good background rejection mechanisms for polarimetry. Using data from the RHESSI satellite an $80\% \pm 20\%$ linear polarization level from GRB 021206 was reported \cite{coburnboggs}, but disproved afterwards \cite{rutledgefox, wigger}. By simulating the scattering of GRB photons off the Earth's atmosphere, the polarization levels of two BATSE bursts (GRB 930131 and GRB 960924) could be obtained \cite{willis} as $>$35\% and $>$50\% respectively, but the result could not be constrained beyond systematics. Also inconclusive were the two measurements \cite{mcglynn, kalemci} obtained using INTEGRAL data from the GRB 041219A. Although they found a high level of polarization ($\sim$60\% and 98\%$\pm$33\%, respectively) they could not statistically claim a polarization detection.

In view of the power and the lack of precise polarization studies of GRBs, several X- and $\gamma$-ray polarimeters have been proposed and are under development. Some examples are POLAR  \cite{produit},  GRAPE (Gamma-Ray Polarimeter Experiment, \cite{bloser}) POET (Polarimeters for Energetic Transients, \cite{hill}), CIPHER (Coded Imager and Polarimeter for High Energy Radiation,  \cite{curado}), PHENEX (Polarimetry for High ENErgy X rays, \cite {gunji}), XPOL \cite{costaXpol} and POLARIX \cite{costaPolarix}. In addition, polarimeters designed for studying fixed sources, like PoGOLite \cite{kamae}, could also measure GRB if they happen to appear in their field of view. 

POLAR \cite{produit} is a small and compact instrument designed to determine the level of linear polarization of the 50 -- 500 keV photons arriving from the prompt emission of GRBs. Totally dedicated to polarimetry, the large field of view ($\sim$1/3 of the sky) of POLAR and its lack of imaging capability would in principle force the detector to rely on other instruments to provide the location of the observed GRBs. Such a limitation would reduce the number of GRBs to be measured by POLAR depending on whether another GRB detector would be observing the same portion of the sky or not. To minimize this drawback we have developed a method to roughly localize GRBs using only POLAR. For this purpose, scalers are attached to the output of the multi-anode photomultipliers (MAPMs) that collect the light from the POLAR scintillator target. Simulations demonstrate that the relative output of those scalers, accumulated over the duration of a GRB, shows a dependency on the source position above POLAR. 

This paper is organized as follows. Section \ref{sec:polar} describes the POLAR detector and its Monte Carlo simulation package. In section~\ref{sec:method} we present the working principle of the GRB localization method in detail. To characterize its capabilities several GRB observations have been simulated. The results obtained are presented in section~\ref{sec:verification}, where the influence on the polarization determination of the uncertainty associated to the estimated GRB position is also quantified. Several systematic effects that could influence the outcome of the method are discussed in section~\ref{sec:systematics}, including GRB polarization, background sources, GRB spectral variations, and satellite backscattering. Finally, section~\ref{sec:summary} summarizes the paper.

\section{Description of POLAR}
\label{sec:polar}
\begin{figure}[h]         	 
\centerline{\includegraphics[width=8cm]{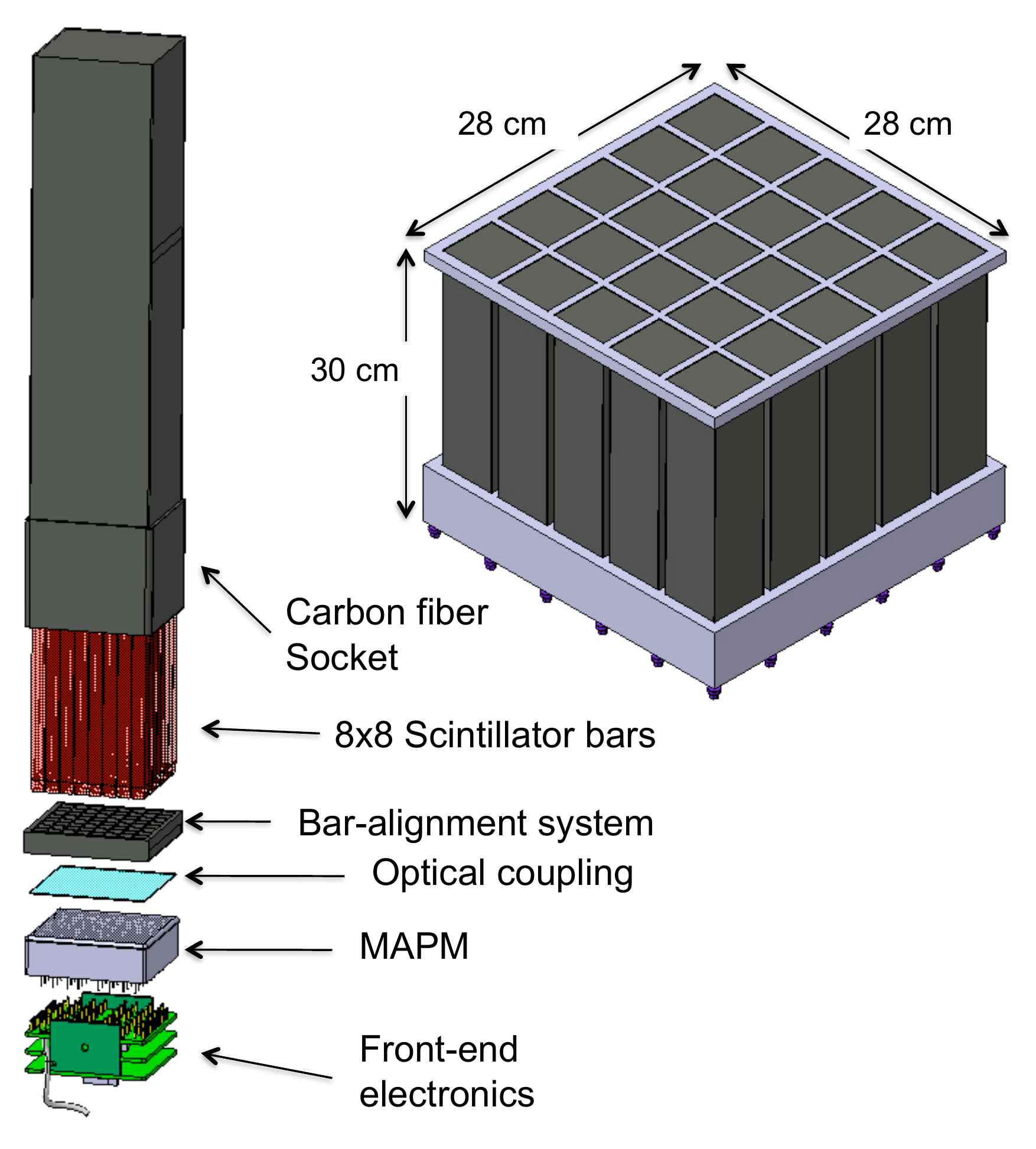}} 
\caption{Scheme of POLAR detector. \textit{Left}: Exploded view of one module from POLAR target. \textit{Right}: Complete POLAR target, i.e., the assembly of 25 modules, with its approximate dimensions.}
\label{fig:1}
\end{figure}

POLAR, see figure~\ref{fig:1}, consists of a target of 40 $\times$ 40 plastic scintillator bars (BC400, Saint-Gobain), each one 6$\times$6$\times$200 mm$^3$ wrapped in a highly reflective foil (Vikuiti, 3M). The optical photons produced when a hard X-ray photon interacts with the target are collected using 25 MAPMs (H8500, Hamamatsu) whose channels are optically coupled to the bottom of the scintillator bars. The electrical signal coming from the MAPM is then processed by ASIC and FPGA at the front-end electronics, and collected by the POLAR central computer which is responsible for data storage and transport to ground. The scintillator target is divided into 25 modules, each consisting of 64 scintillator bars, optical coupling, one MAPM, and its corresponding front-end electronics, all together enclosed in a thin carbon fiber socket. This modular design gives a good mechanical stability and facilitates the interchange of modules during the testing phase of the detector. The 25 modules are kept together with two aligning frames located at the top and at the bottom of the carbon fiber sockets, respectively and connected with mechanical fixations. The whole target, together with the central computer, the power supplies and the rest of the electronics, is further enclosed in a box that serves not only as container but also as shield against low energy charged particles. Finally, the whole instrument will be mounted onto a satellite, to be able to study photons in the energy range between 50 and 500 keV, which cannot reach the ground because they are absorbed by the Earth atmosphere. A flight opportunity for POLAR on the future Chinese Tian-Gong Space Station is currently under consideration.  An alternative option of flying with the International Space Station is evaluated.

An incoming particle is called an \textit{event}  when it deposits more than 5 keV in at least one of the POLAR scintillator bars, i.e when it produces a \textit{hit}. Hard X-ray photons arriving from a GRB have a high probability of experiencing Compton scattering in the target, generating a signal in more than one channel. The POLAR trigger logic selects those events with at least two hits. When an incoming particle produces more than two hits above the 5 keV threshold, the two highest energy depositions are selected offline. Each of those pairs of hits is a \textit{selected event} in POLAR. The position of the active bars in a selected event is related to the azimuthal Compton-scattering angle of the incoming photon. If the $\gamma$-ray emission from the GRB was not polarized, the modulation curve, i.e. the azimuthal distribution of the ensemble of selected events, is flat. Otherwise it follows a sinusoidal curve whose amplitude is the so-called \textit{modulation factor}, and whose phase-shift indicates the angle of linear polarization of the GRB photons \cite{mcconnell, lei}. The modulation factor ($\mu$), divided by the response of the detector to a 100\% polarized flux ($\mu_{100}$), is the polarization level ($\Pi$) of the incoming photons:  $\Pi = \mu/\mu_{100}$.  This principle allows POLAR to measure the level of polarization of the prompt emission of GRBs.

The main part of the POLAR front-end electronics is devoted to collect the signals produced in the 1600 MAPM anodes, to implement the trigger, and to transport the triggered signal to the central computer. The trigger condition is fulfilled when the target registers at least two hits above the 5 keV single channel threshold in a time window of $\sim$70 ns. Only the events with this characteristic will arrive to the central computer and be available for analysis. The probability of registering an accidental coincidence is negligible considering the counting rates expected during a GRB.  

For the GRB localization method, the electronics need to include in addition 25 scalers, one per MAPM, counting all photons that pass a higher energy threshold. If the final electronics design includes a 2-level ASIC, it is possible to define a low energy single channel threshold of 5 keV to use in the trigger logic, and a higher energy threshold of 50 keV per channel, useful for the GRB localization method implementation. A simple "or" operation on the signals that have passed the high level threshold, during a time span of about 70 ns, would be enough to know if at least one hit with more than 50 keV energy deposition has been produced in the corresponding MAPM. An alternative option for the case where a single level ASIC is used would be to attach the scaler to the dynode output of each MAPM, that provides an analog sum of the 64 channels. All the results presented in this paper have been obtained with Monte Carlo simulations that reproduced the first approach. Nevertheless, independently of which approach is taken, the final scaler output for one event is a collection of 25 numbers with values 0 or 1. When accumulating the scaler output for all events produced during a GRB one obtains the number of photons that interacted above the given high energy threshold inside every MAPM.  The counting pattern is sensitive to the direction of incoming photons since the size of the target corresponds to several absorption lengths. The scalers cannot distinguish between GRB photons and background. For this reason the influence of the background has been studied and is presented in section~\ref{sec:influencebg}. The algorithm is relatively simple and suitable to be implemented in a FPGA or in the small on board processor. The needed computing power is negligible as the algorithm has to run only infrequently when a candidate GRB has been detected.

An important part of the POLAR design and the evaluation of its polarimetric capabilities is performed using GEANT4 \cite{geant4} simulations. The 1600 scintillator bars of the target have been simulated with their standard dimensions and wrapped with a 50 $\mu$m thick aluminum foil. The target is further placed inside an aluminum box of 1 mm thickness that serves as a shield from low-energy charged particles and represents the outer enclosure of the detector. The incoming direction, the spectrum, and the type of particles to be used are generated following the user instructions. A specific routine has been developed to produce hard X-ray photons following the Band model spectrum \cite{band} and its parameters. The most important physical processes that photons can undergo in the detector are taken into account, including polarized Compton scattering, photoelectric effect, and pair production, among others. The scintillation and optical light collection processes have not been considered. The simulation generates a ROOT \cite{brun} file containing all the necessary data for the subsequent analysis: the incoming photon energy, the number and position of the bars fired, the energy deposited at each bar, etc.

\section{GRB Localization Method}
\label{sec:method}

The pattern of the scalers output (SO) presented in section~\ref{sec:polar} depends on the position of the GRB above the target. A GRB located at POLAR zenith will produce an SO with more counts in the entries corresponding to the central MAPMs. On the other hand, a GRB at a large polar angle will present more counts in the MAPMs more directly illuminated by the GRB because of the exponential interaction probability. Using this relation it is possible to localize the sources observed by POLAR.

For this purpose a database of SOs has been created using the POLAR Monte Carlo simulation package. Non-polarized GRBs have been simulated at numerous incoming angles. To get enough statistics for the comparison with future GRB measurements, we simulated very strong bursts for the database, with a total energy fluence $\rm{F_{tot}} = 10^{-4}$ erg cm$^{-2}$ and Band spectral parameters $\alpha=-1.0$, $\beta=-2.5$, and $\rm{E_{peak}}=200$ keV \cite{band}. The selected Band parameters correspond to the approximate average values calculated with a selection of strong GRBs observed by BATSE \cite{preece, kaneko}. 

Defining in spherical coordinates the positions of the simulated GRBs in the sky would imply an indetermination of the incoming photon azimuthal angle ($\phi_\gamma$) when its polar angle ($\theta_\gamma$) approaches 0. To avoid this numerical problem the database has been produced in Cartesian coordinates:

\begin{equation}     
\begin{array}{ll}
 x = \left(1-\cos\theta_{\gamma} \right) \cdot \cos\phi_{\gamma} \\	
 y = \left(1-\cos\theta_{\gamma} \right) \cdot \sin\phi_{\gamma} \\ 	
\end{array} 
\label{eq:1}
\end{equation}

Simulations were uniformly distributed in the ($x$,$y$) plane, for $x$ and $y$ running between -1 and 1 in steps of 0.02. We obtain in this way a grid with 10201 nodes. Note that, since the detector is symmetric and the influence of the satellite behind it is negligible (see section~\ref{sec:satellite}), it is enough to simulate one quadrant of the grid. The SOs of the other three quadrants can be inferred from the 
simulated ones. The selection of the coordinate system from equation~\ref{eq:1} guarantees a uniform grid where all points are equally distant from each other and the center of the coordinate system corresponds to the zenith of the detector. 

\begin{figure}[h]        	 
\centerline{\includegraphics[width=4cm]{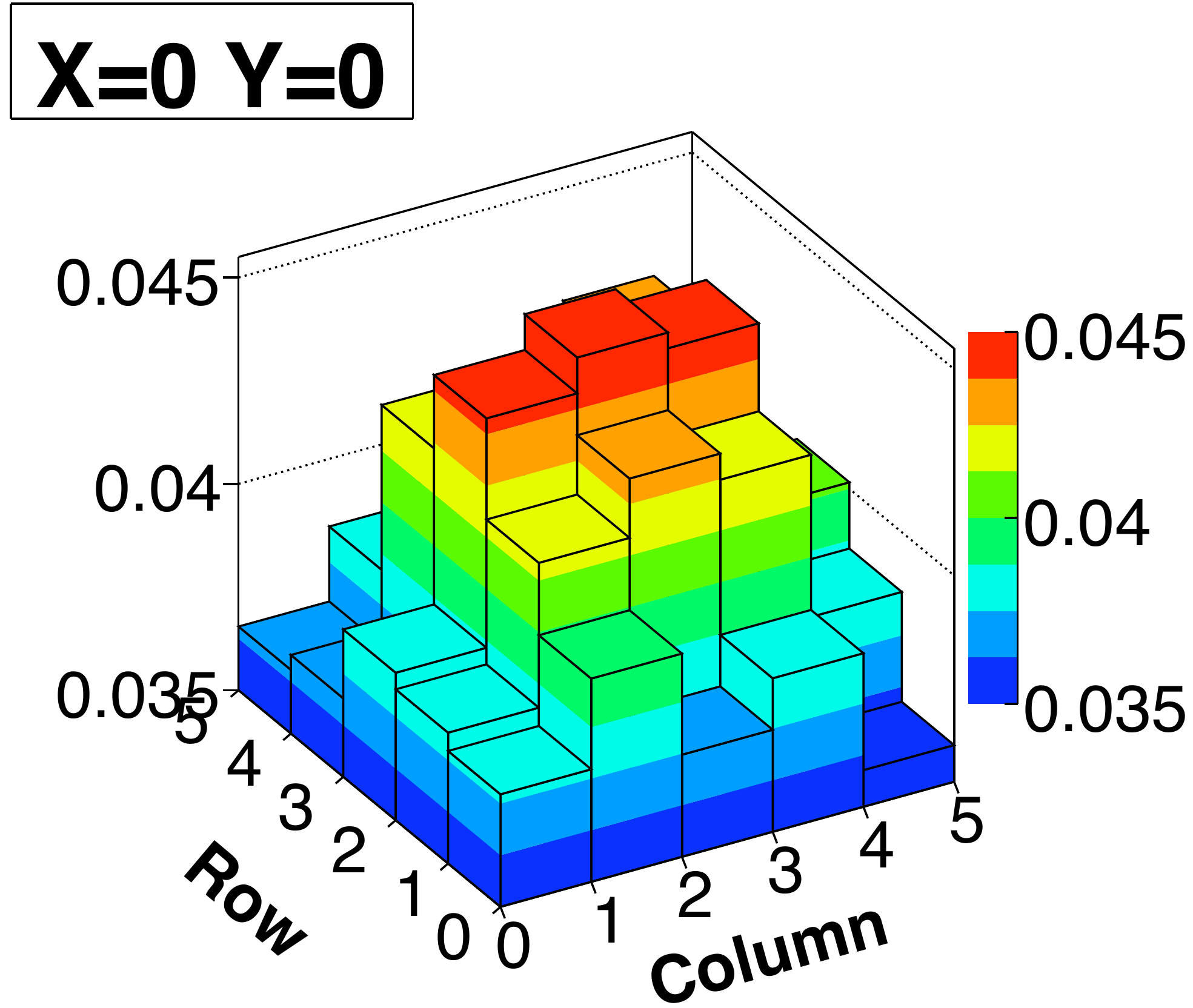} \includegraphics[width=4cm]{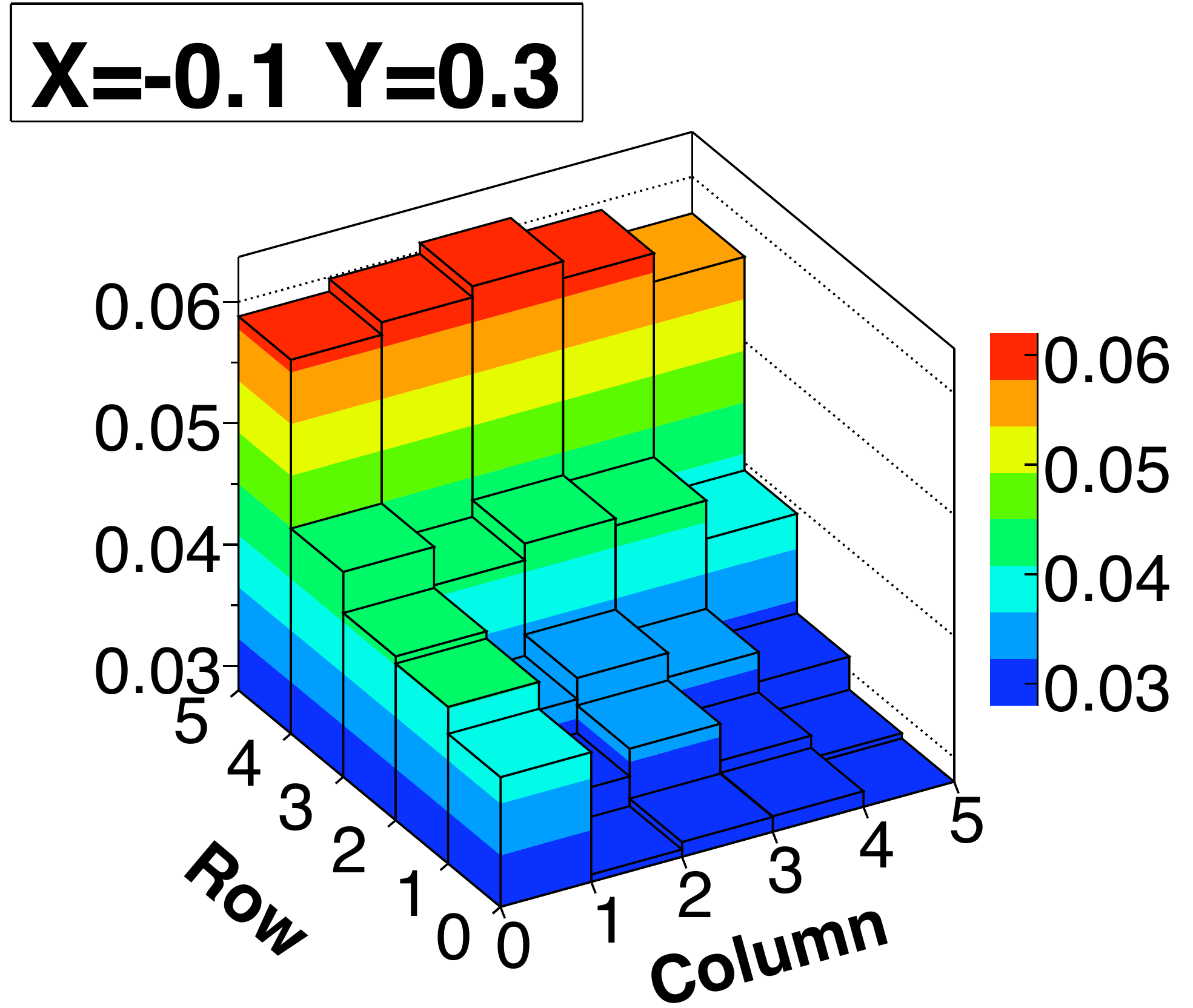} }
\caption{Graphical representation of two of the normalized scaler output (NSO) patterns that constitute the database.}
\label{fig:2}
\end{figure}

The output of the scalers was calculated considering only hits with more than 50 keV deposited energy, following the first approach presented in section~\ref{sec:polar}. Although there are no intrinsic limitations that prevent the detection of weaker energy depositions, the inclusion of this threshold is necessary to reduce the spectral dependence of our localization method (see section~\ref{sec:influencespec} for details). Furthermore, a normalization of the simulated SOs is needed due to the dependency of the absolute number of counts registered in each MAPM on the total flux that the source produced. We scaled the rates dividing by the sum of all counts in the given SO. The result is a normalized scaler output (NSO) like the ones graphically represented in figure~\ref{fig:2}. The final database is constituted by 10201 text files containing the NSO and its statistical error for each of the 25 MAPM of POLAR target.

The process to localize the GRBs detected by POLAR is based on the minimization of the Pearson's $\chi^2$ \cite{hauschild, baker}:
\
\begin{equation}  		
\chi^2=\sum^{n}_{i=1} \frac{\left( c_i - m_i \right)^2}{m_i},
\label{eq:2}
\end{equation}

\noindent where $n$ is the number of MAPMs, $c_i$ is the number of counts observed in MAPM $i$, and $m_i$ is the number of counts predicted by a given model for the same MAPM $i$. When POLAR observes a GRB, we calculate $c_i$ as the $i^{th}$ entry of its measured SO, and $m_i$ is obtained from the database of simulated NSOs. Supposing that the GRB was at coordinates ($x$,$y$), the predicted number of counts in the $i^{th}$ MAPM is the $i^{th}$ entry of the database file at this position of the sky ($m_i(x,y)$) multiplied by the total number of entries ($c_{\rm tot}$) in the measured SO:

\begin{equation}  		
\chi^2{(x,y)}=\sum^{25}_{i=1} \frac{\left( c_i - c_{\rm tot}  \cdot m_i(x,y) \right)^2}{c_{\rm tot}  \cdot m_i(x,y)}.
\label{eq:3}
\end{equation}

\begin{figure}[h]         	 
\centerline{\includegraphics[width=8cm]{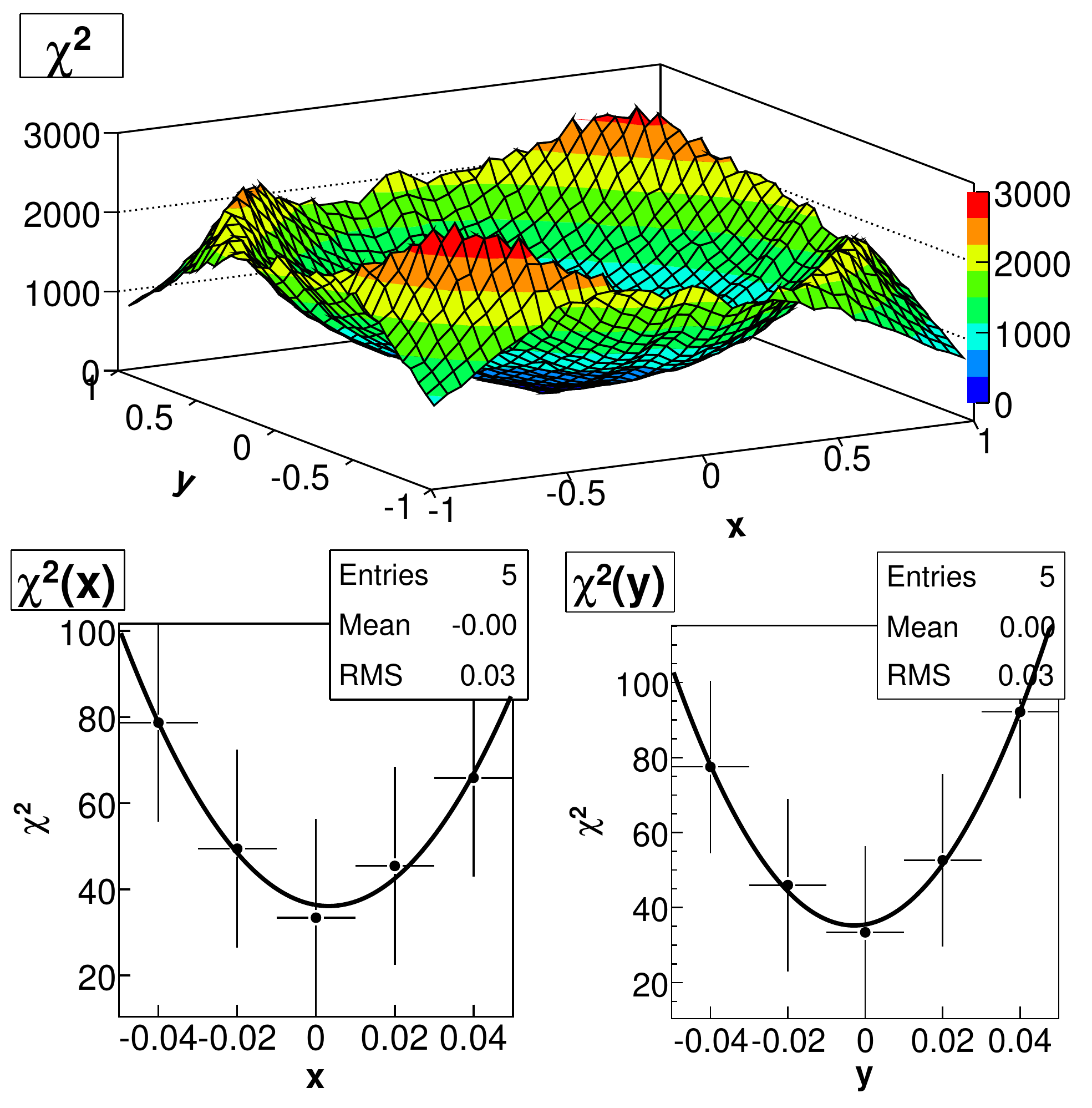}} 
\caption{Example of $\chi^2$ minimization. The distribution of  $\chi^2$ in the ($x$,$y$) plane is shown in the top, and the parabolic fits of both variables in the bottom. The measurement used in this example was produced using the POLAR simulation package with a non-polarized GRB of F$_{\rm tot} = 10^{-5}$ erg cm$^{-2}$, $\alpha$=-1.0, $\beta$=-2.5, $E_{\rm peak}$=200 keV located at POLAR zenith.}
\label{fig:3}
\end{figure}

To localize a GRB we calculate all of 10201 $\chi^2$ values, one at each point of the database grid, obtaining a 2-dimensional  $\chi^2$ distribution (see upper part of figure~\ref{fig:3} for an example done with a burst of same spectral shape as the database, but fluence F$_{\rm tot} = 10^{-5}$ erg cm$^{-2}$). The minimum of the distribution is found selecting the node of the grid with the lowest $\chi^2$ and fitting two parabolas around it, one in the $x$-direction and the other in the $y$-direction, using 5 points in each case. The grid is fine enough so that the result obtained in this way is a good approximation to the absolute minimum of the two-dimensional distribution. To perform the parabolic fits with ROOT one needs to assign an error to each of the points. Since the $\chi^2$ itself has no associated error we assigned a fixed value, equal for all the points. We have chosen this value equal to 23, which is the number of degrees of freedom of our experiment. On the horizontal axis the error of each point is the half-width of the bin, in our case 0.01. The positioning of the minimum of the parabolas $\chi^2_{\rm min}(x)$ and $\chi^2_{\rm min}(y)$ corresponds to the estimated position of the source. The error on the estimation is calculated with the values of the abscissa located at $\chi^2_{\rm min}(x) + 1$ and $\chi^2_{\rm min}(y) + 1$ respectively. Figure \ref{fig:3} shows an example of the two-dimensional distribution of $\chi^2$ values, and the two fitted parabolas.   

In figure~\ref{fig:3} one can notice that the $\chi^2$ values do not grow monotonously from the minimum, but decrease at the four corners of the plane. These four corners of the Cartesian space correspond to situations where the GRB is below the POLAR detector, i.e. $\theta_\gamma>90^\circ$. The $\chi^2$ decreases there because POLAR cannot measure the point of interaction of the photons along the length of its bars, and is therefore incapable of distinguishing a GRB at ($\theta_\gamma$, $\phi_{\gamma}$) from one at (180$^\circ$-$\theta_{\gamma}$, $\phi_{\gamma}$). To avoid this ambiguity, and since GRBs will never appear below POLAR, we have introduced an extra condition in the localization method that takes it into account. Namely, every time that the method gives an output below POLAR it will be transformed into the equivalent position above POLAR and only this one will be given as final result. In practice one would actually not need to simulate those points of the database. Our reason to do it is the numerical advantage of working with a uniform and square grid that can be treated as a regular matrix.

\section{Method verification}
\label{sec:verification}

\begin{figure*}[!ht]         	 
\centerline{\includegraphics[width=14cm]{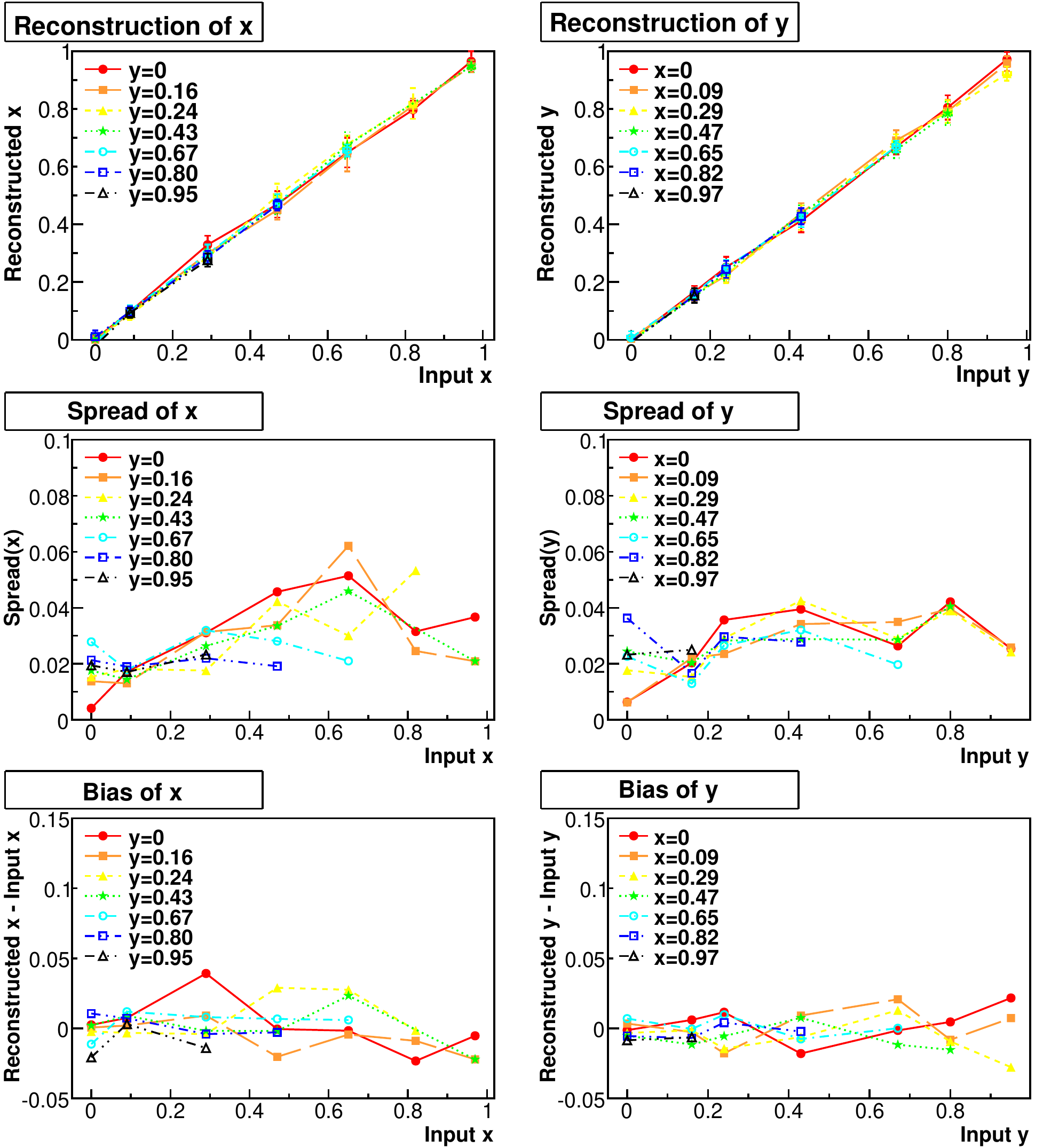}} 
\caption{Results from simulated non-polarized \textit{standard} strong GRBs. \textit{Top left}: Linearity plot in $x$,  i.e. reconstructed $x$ vs. input $x$ for several fixed $y$ values.  Each point on this graph represents the average of the 10 simulations made at each position. \textit{Middle left}: Spread of the reconstructed $x$, i.e. the standard deviation of the 10 reconstructed $x$ values. \textit{Bottom left}: Bias in the reconstruction, i.e. the difference between the reconstructed average $x$ value and the input $x$. The plots on the right are the equivalent to the left ones, but calculated for $y$ at several fixed $x$ positions. In all the plots lines of different styles have been drawn to guide the eye.}
\label{fig:4}
\end{figure*}

Simulations were made to produce \textit{example measurements} with which we could test the localization method. The GRBs simulated as examples were non-polarized \textit{standard} strong GRBs. Hereafter, \textit{standard} strong stands for a GRB with total energy fluence $\rm{F_{tot}} = 10^{-5}$ erg cm$^{-2}$ and Band spectral parameters \cite{band} $\alpha=-1.0$, $\beta=-2.5$, and $\rm{E_{peak}}=200$ keV, from  which POLAR will detect about 22000 photons in the whole duration of the burst. Note that the spectal shape, i.e. the Band parameters, is the same as used for the database bursts, but the total fluence is here ten times smaller. We expect, considering the BATSE catalog and the POLAR field of view, to observe about 12 such \textit{standard} strong GRBs per year \cite{hajdas}. To facilitate the visualization we present only GRB positions above POLAR and in the quadrant of the space where $x>0$ and $y>0$. The behavior on the other three quadrants is equivalent due to the symmetry of the detector. 

We compared the $x$ and $y$ values taken as input in the \textit{measurements} with the ones reconstructed using the $\chi^2$ minimization method. Ten samples were run at each selected ($x$,$y$) position to study the repeatability and the spread of the reconstructed results around each input value. Figure \ref{fig:4} shows the performance of the GRB localization method. The localization procedure works well reconstructing  both $x$ and $y$ without apparent bias within a root-mean-square (r.m.s.) smaller than 0.07. The average minimum $\chi^2$ obtained for the 380 measurements simulated to produce figure~\ref{fig:4} was $\chi^2_{\rm min}(x,y) = 22.9\pm6.6$. Taking into account that there are 23 degrees of freedom, this corresponds to a reduced $\chi^2_{red}(x,y) = 0.996 \pm 0.287$.  When performing the equivalent GRB localization in spherical coordinates $\theta_\gamma$ could be determined within a r.m.s  $\approx4^\circ$ and $\phi_\gamma$ within a r.m.s $\approx5^\circ$ for $\theta_\gamma > 30^\circ$. As comparison, the localization accuracy of the Fermi-GRB burst monitor (GBM) \cite{lichti} is for bright bursts $<1.5^\circ$ using a detailed offline analysis, and the Swift Burst Alert Telescope (BAT) \cite{barthelmy} provides positions to an accuracy of 1' -- 4' within 20 seconds of GRB observation.


When a GRB is detected by POLAR its level of polarization ($\Pi$) is determined from the amplitude ($\mu$) of the modulation curve and the 100\% modulation factor ($\mu_{100}$) using:  $\Pi = \mu/\mu_{100}$ (see section~\ref{sec:polar}). The error on the polarization level can be calculated as:

\begin{equation}  		
\frac{\sigma_{\Pi}}{\Pi}= \sqrt{\left( \frac{\sigma_{\mu_{100}}}{\mu_{100}} \right)^2 + 
	\left( \frac{\sigma_{\mu}}{\mu} \right)^2},
\label{eq:4}
\end{equation}

\noindent where  $\sigma_{\Pi}$, $\sigma_{\mu}$ and $\sigma_{\mu_{100}}$ are the errors on $\Pi$, $\mu$ and $\mu_{100}$, respectively.

\begin{figure}[!h]         	 
\centerline{\includegraphics[width=8cm]{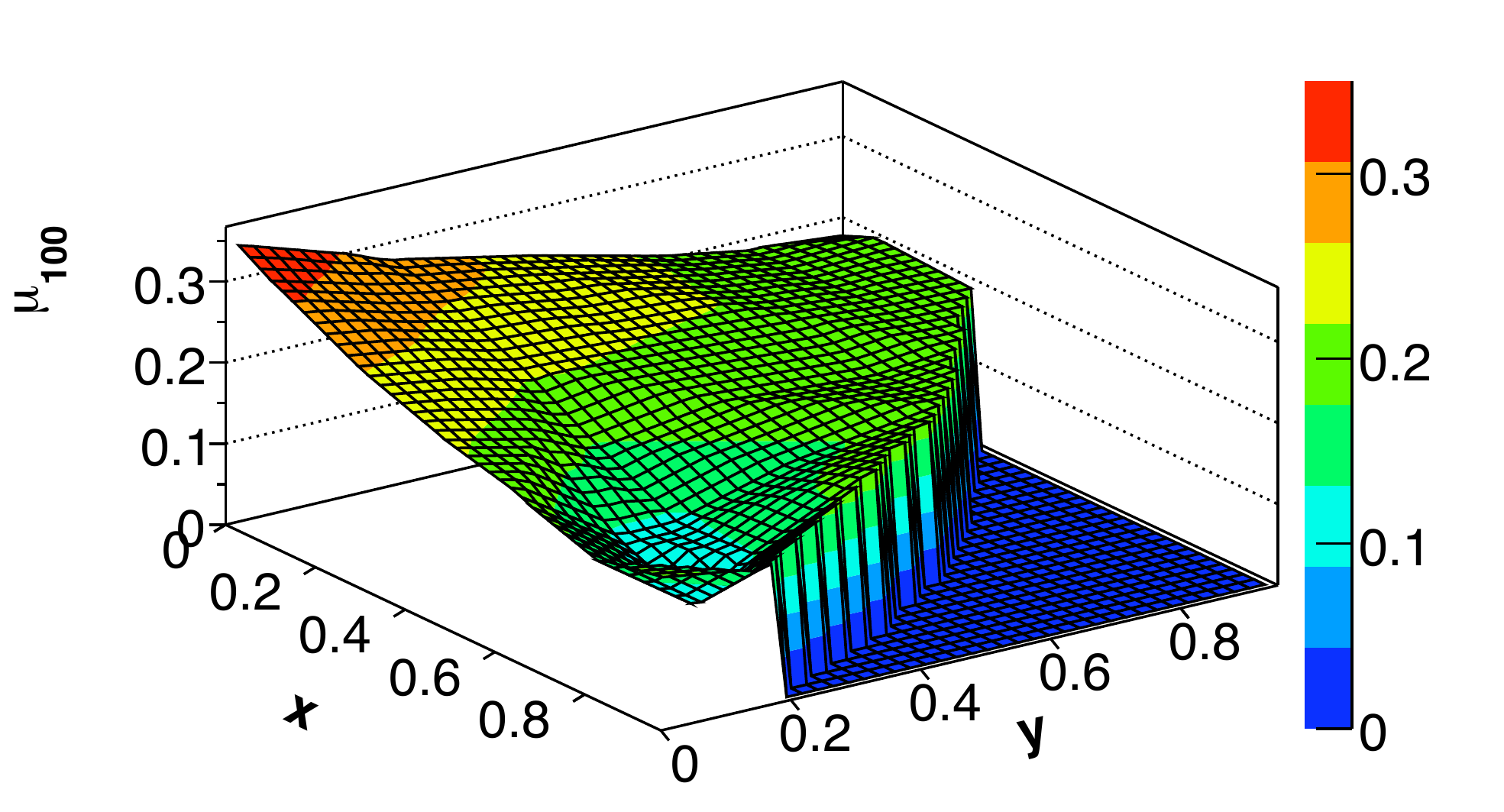}} 
\caption{Distribution of 100\% modulation factor ($\mu_{100}$) vs. $x$ and $y$, calculated for a standard strong GRB (see text). The empty area at the right of the image corresponds to positions where the GRB would be below POLAR. Such a situation will never happen during flight and therefore was not considered in the analysis.}
\label{fig:5}
\end{figure}

The second term in  equation~\ref{eq:4} depends only on the strength and level of polarization of the observed GRB. Since $\mu_{100}$ varies with the burst position and spectral shape, $\Pi$ must be calculated using a $\mu_{100}$ derived from simulations of a GRB with the same spectrum and position in the sky as the observed one. The highest $\mu_{100}$ is obtained when the GRB is located at POLAR zenith and it diminishes when photons come from POLAR side (see figure~\ref{fig:5}). When the location of the GRB  is unknown, the value of $\mu_{100}$ needed to calculate $\Pi$ is also unknown. Therefore the uncertainty on the position of the GRB is translated to an uncertainty in the value of  $\mu_{100}$ and, through the first term of equation~\ref{eq:4}, further into an error on the resultant level of polarization.

One can calculate the error of  $\mu_{100}$ with standard error propagation:

\begin{equation}  		
\sigma_{\mu_{100}}= \sqrt{\left( \frac{\partial \mu_{100}}{\partial x}\sigma_x \right)^2 + 
 	\left( \frac{\partial \mu_{100}}{\partial y}\sigma_y \right)^2}
\label{eq:5}
\end{equation}

For each of the positions simulated as examples, the spread and bias of the reconstructed values were determined. When the spread of the reconstructed coordinates was larger than their bias, the spread was introduced as  $\sigma_x$ and $\sigma_y$ in equation~\ref{eq:5}. Otherwise, the value of the bias was taken. To calculate the partial derivatives of the $\mu_{100}$ with respect to $x$ and $y$ we fitted the 2-dimensional distribution from figure~\ref{fig:5} with the following polynomial function:

\begin{equation}  		
\mu_{100}(x,y)= a_0+a_1x+a_2x^2+a_3y+a_4y^2+a_5xy
\label{eq:6}
\end{equation}

\noindent whose partial derivatives are:


\begin{equation}  		
 \left\{ 
 \begin{array}{ll}
  \frac{\partial \mu_{100}}{\partial x} =  & a_1 + 2a_2x + a_5y \\
  & \\
  \frac{\partial \mu_{100}}{\partial y} =  & a_3y + 2a_4y + a_5x \\ 
\end{array} 
\right. 
\label{eq:7}
\end{equation}

\begin{figure}[h]         	 
\centerline{\includegraphics[width=8cm]{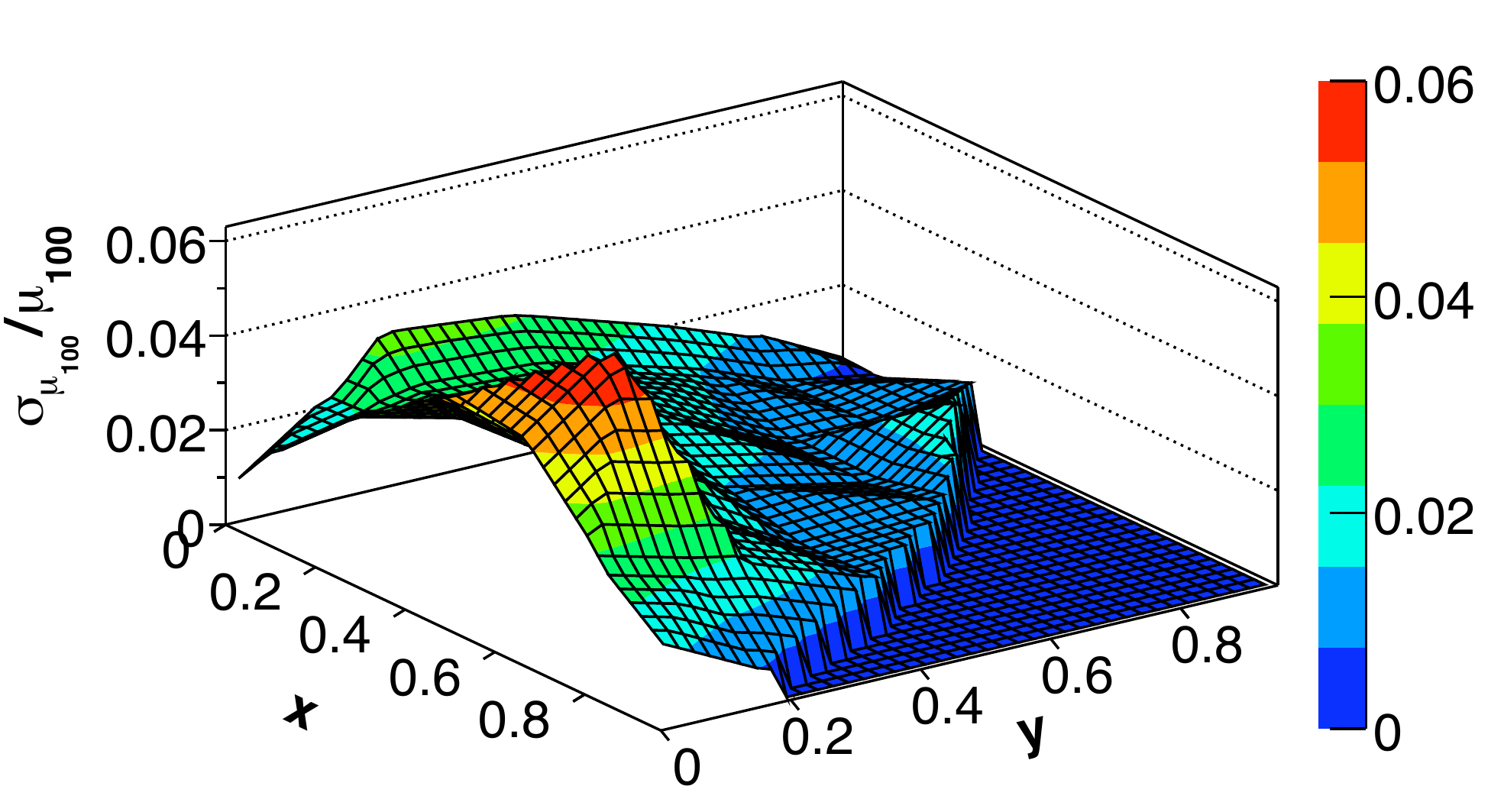}} 
\caption{Error caused to $\mu_{100}$ due to the uncertainty on the localization of GRBs, divided by $\mu_{100}$.}
\label{fig:6}
\end{figure}

From equations \ref{eq:7} and \ref{eq:5} one can calculate the error induced on $\mu_{100}$ by the uncertainty in the location of the GRB. The final result can be seen in figure~\ref{fig:6} and represents the value of the first term in equation~\ref{eq:4} $(\sigma_{\mu_{100}} / \mu_{100})$ at various source locations. The error caused to $\mu_{100}$ is always smaller than 6\% of its absolute value. 

\begin{figure}[!h]         	 
\centerline{\includegraphics[width=4cm]{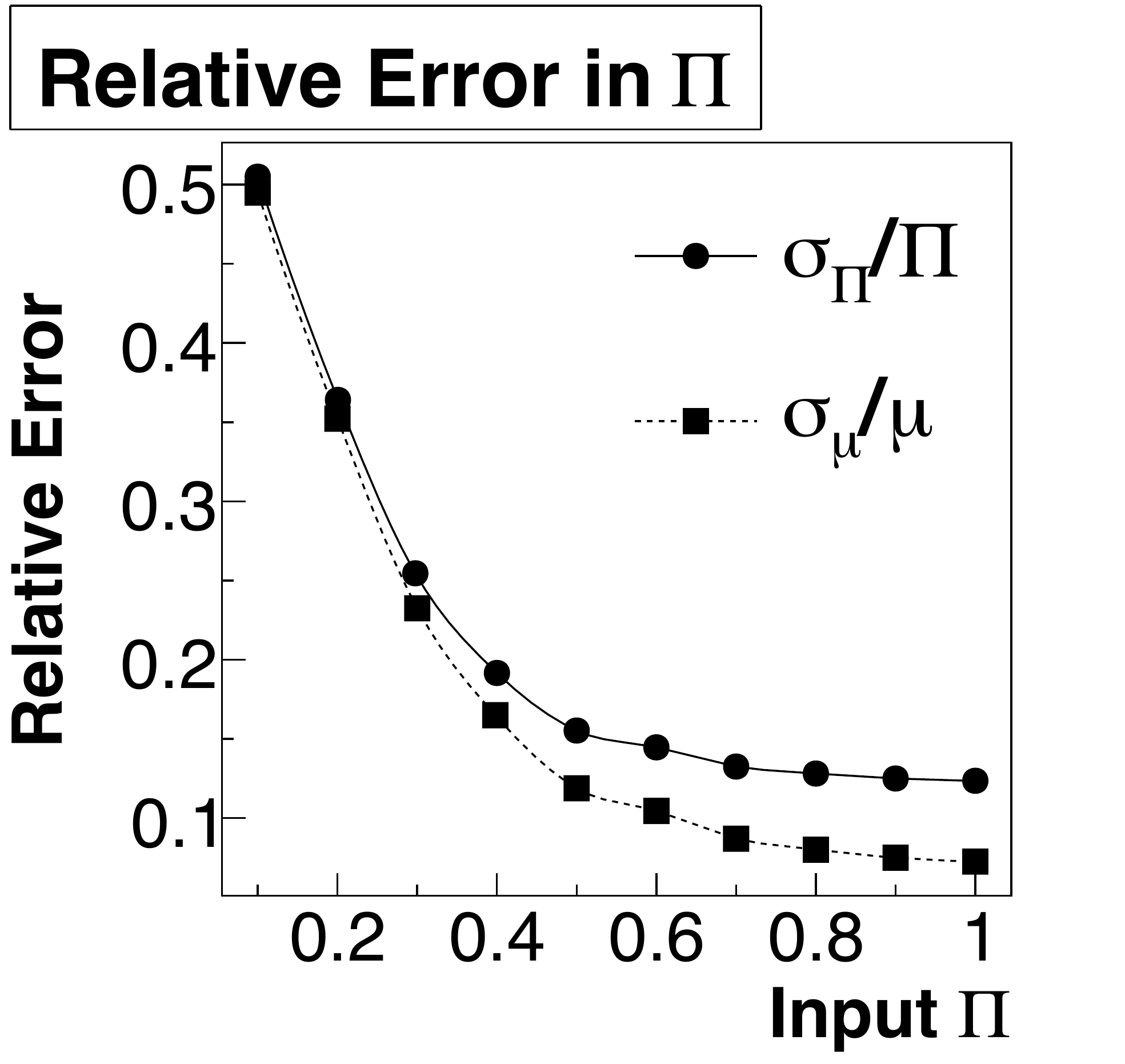} \includegraphics[width=4cm]{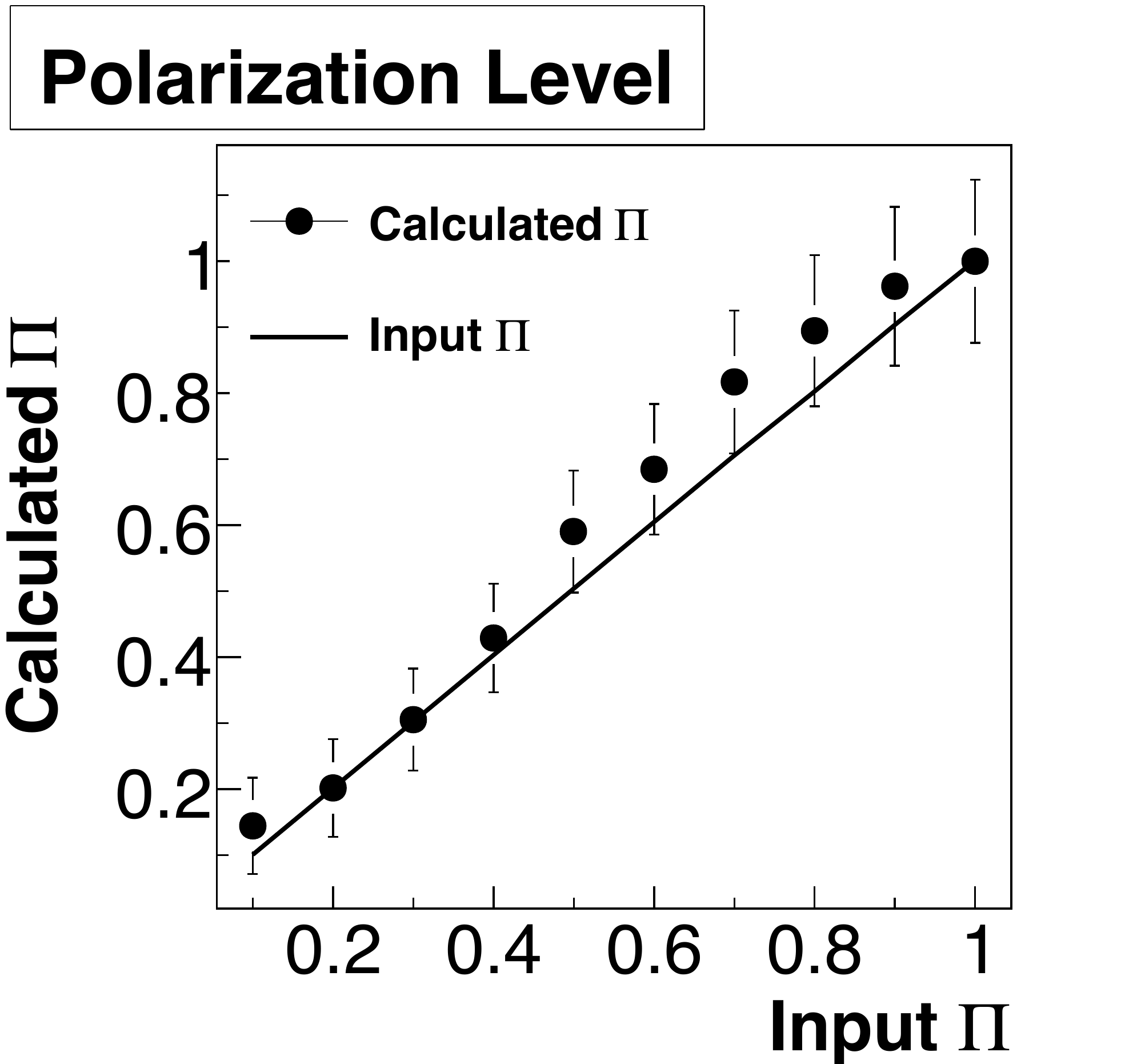}} 
\caption{Upper limits to the error on the polarization level for a GRB of total fluence ${\rm F_{tot}}=10^{-5}$ erg cm$^{-2}$.  These plots have been calculated using the upper limit of the error inflicted to $\mu_{100}$ due to the localization procedure.  {\it Left}: Relative error of the modulation factor for different levels of polarization ({\it dotted}), together with the final relative error on $\Pi$  ({\it solid}), where $\sigma_{\mu_{100}}/\mu_{100}=0.1$ has been included. The increase due to the localization uncertainty is visible at high polarization values. {\it Right}: Calculated polarization level ({\it points}), together with the simulation input values ({\it line}), plotted as a reference. For polarization levels above 60\% the uncertainty in the GRB position is the largest source of error.}
\label{fig:7}
\end{figure}

The total error in the measured polarization is given by equation~\ref{eq:4}. Its second term ($\sigma_\mu / \mu$) depends on the strength and level of polarization of the GRB. In figure~\ref{fig:7} we present an example calculated for \textit{standard} strong bursts of fluence ${\rm F_{tot}}=10^{-5}$ erg cm$^{-2}$, located at POLAR zenith, and with various levels of polarization. For this calculation we have substituted $\sigma_{\mu_{100}}/\mu_{100}=0.10$ in equation~\ref{eq:4}, instead of the 0.06 previously mentioned, because 10\% was the maximal error found on $\mu_{100}$ for such a \textit{standard} strong GRB when variations of the spectral shape were taken into account (see section~\ref{sec:influencespec}). One can notice that the error due to the position uncertainty is negligible for low levels of polarization, with respect to the error associated to the modulation factor itself. At large levels of polarization the localization contribution to the error is more important. The final error in $\Pi$ is in that case $\sigma_\Pi \le15\%$ of the absolute polarization level. The values represented in figure~\ref{fig:7} should be considered for bursts of total fluence ${\rm F_{tot}}=10^{-5}$ erg cm$^{-2}$ as upper limits to the error on the polarization level.

\section{Systematic effects}
\label{sec:systematics}
The results presented in figure~\ref{fig:4} represent the best case example, since the measurements to which the method was applied were produced in the same conditions as the localization database, and considering POLAR as an instrument with a perfectly uniform response. During a real observation there are several issues that could influence the outcome of the GRB localization method. We will discuss here the most important ones: asymmetries produced by GRB polarization, diffuse photon background, GRB spectral variations, satellite backscattering, fluctuations in light-collection efficiency, and MAPM non-uniform sensitivity.

\subsection{GRB polarization}
The database of the localization method was constructed simulating unpolarized sources. Although the polarization introduces an asymmetry in the POLAR hit pattern, it does not strongly affect the output of the scaler used for the localization procedure. Some examples were made with simulations of \textit{standard} strong GRBs, 100\% polarized in the direction parallel to POLAR X-axis, to confirm that statement. As with the non-polarized examples, 10 simulations were produced at each position of the sky and the spread and bias in the reconstruction of their locations were calculated. It was clearly seen that the outcome of the localization procedure was not affected by the polarization of the source. Both the spread of the reconstructed values and the bias of their average were within the errors equal to the non-polarized case.

\subsection{Diffuse $\gamma$-ray background}
\label{sec:influencebg}
Outside of the South Atlantic Anomaly the largest source of background that will affect POLAR on orbit is the diffuse $\gamma$-ray photon background \cite{produit}. This source of background illuminates approximately isotropically the POLAR instrument. Since the surface of the most external MAPMs is more exposed, the SO produced by the $\gamma$-ray diffuse photon background presents higher number of counts for the external MAPMs and lower in the internal ones. When analyzing the SO of the total signal, i.e. GRB+background, the result of the localization method may be different than the obtained from the GRB alone. POLAR will store data from before, during, and after the GRB, making possible the subtraction of the diffuse $\gamma$-ray background. We will discuss here how accurately the interpolation and subtraction need to be made so that the localization method is not affected.

The spectrum of the $\gamma$-ray diffuse photon background has been parametrized \cite{gruber} using the function:

\begin{equation} 		
\begin{array}{rll}
\mbox{ 3--}  60 \mbox{ keV} &:& \\
 f(E) & = & 7.877 E^{-0.29} e^{-E/41.13} \\ \\
> 60 \mbox{ keV} &:&  \\
 f(E) & = & 0.0259 \left( \frac{E}{60}\right)^{-5.5} + \\
& & + 0.504 \left( \frac{E}{60}\right)^{-1.58}  + \\
& & + 0.0288 \left( \frac{E}{60}\right)^{-1.05} 
\end{array} 
\label{eq:8}
\end{equation}

\noindent with $f(E)$ in units of  $\frac{\mbox{keV}}{\mbox{keV cm}^2 \mbox{ s sr}}$. An increment of 10\% in the absolute normalization factor has been needed to fit recent measurements of the cosmic hard X-ray spectrum performed by INTEGRAL instruments \cite{churazov}.

\begin{figure}[!h]         	 
\centerline{\includegraphics[width=4cm]{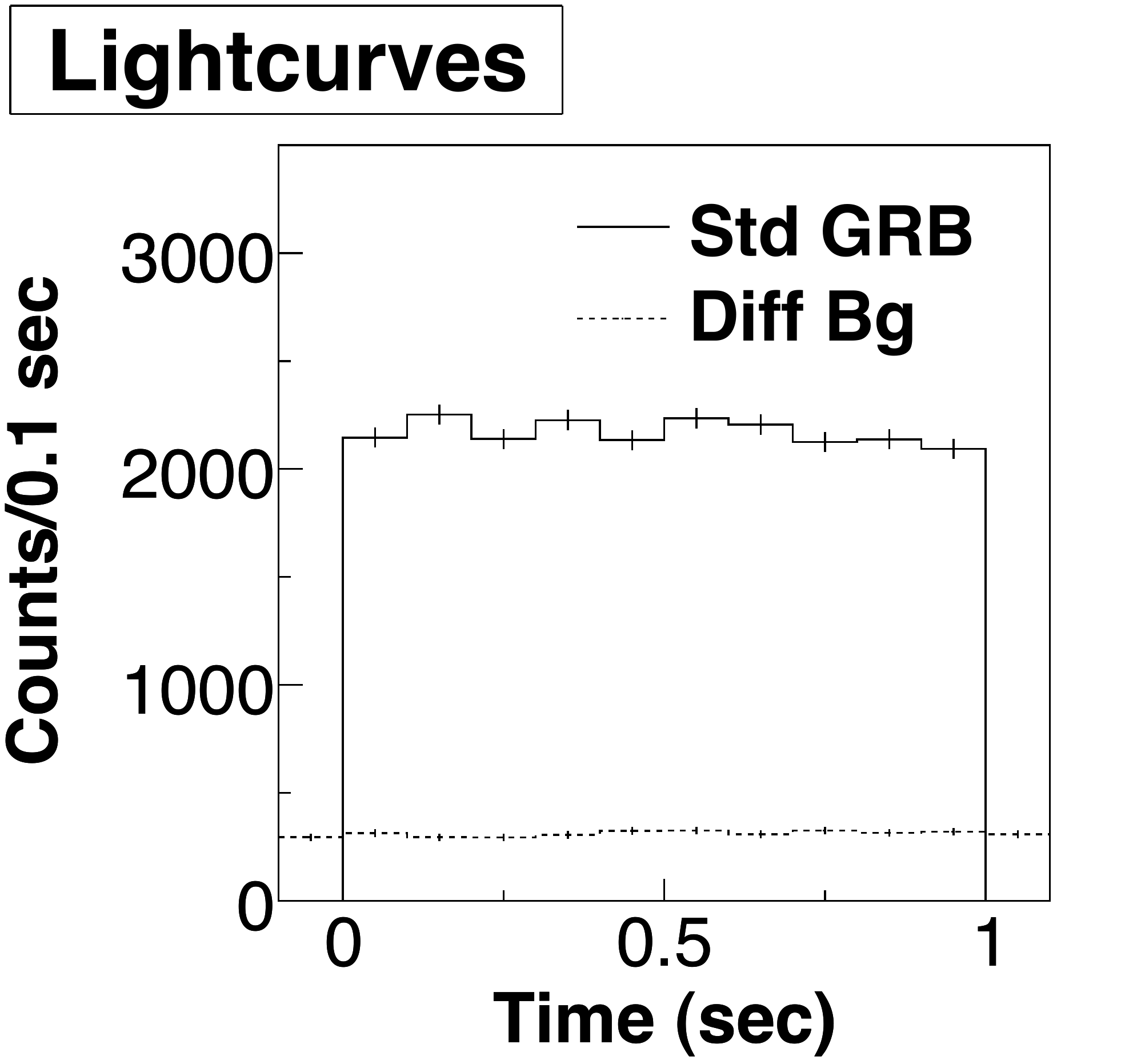} \includegraphics[width=4cm]{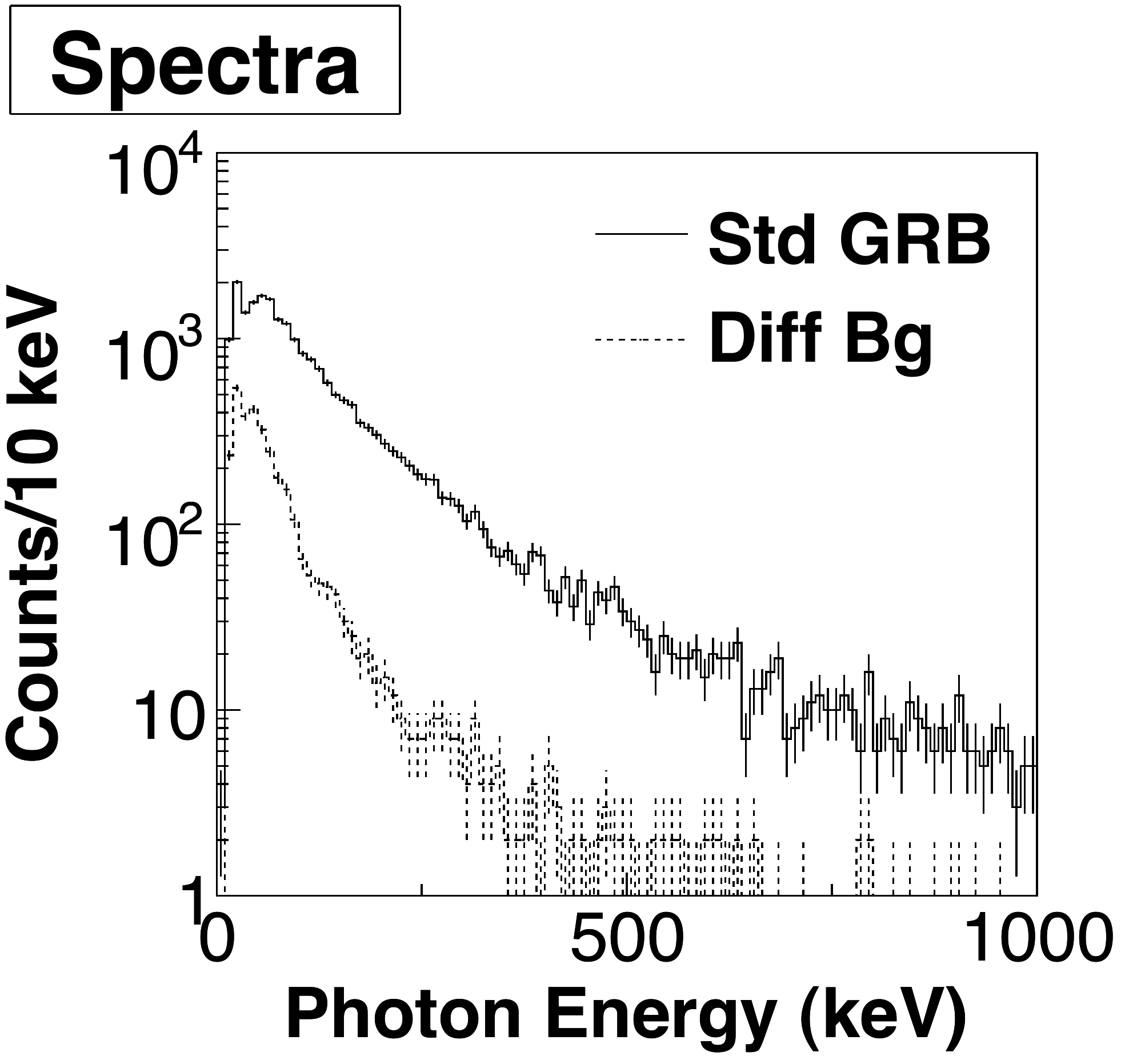}} 
\caption{Comparison between the flux of a short (1 sec) \textit{standard} strong GRB and the diffuse photon background. {\it Left}: Assumed lightcurve of the GRB signal ({\it solid}) and the diffuse photon background ({\it dotted}). All photons producing a hit above 5 keV in POLAR target have been included. {\it Right}: Spectrum for the same two sources.}
\label{fig:8}
\end{figure}

\begin{figure}[!h]         	 
\centerline{\includegraphics[width=4cm]{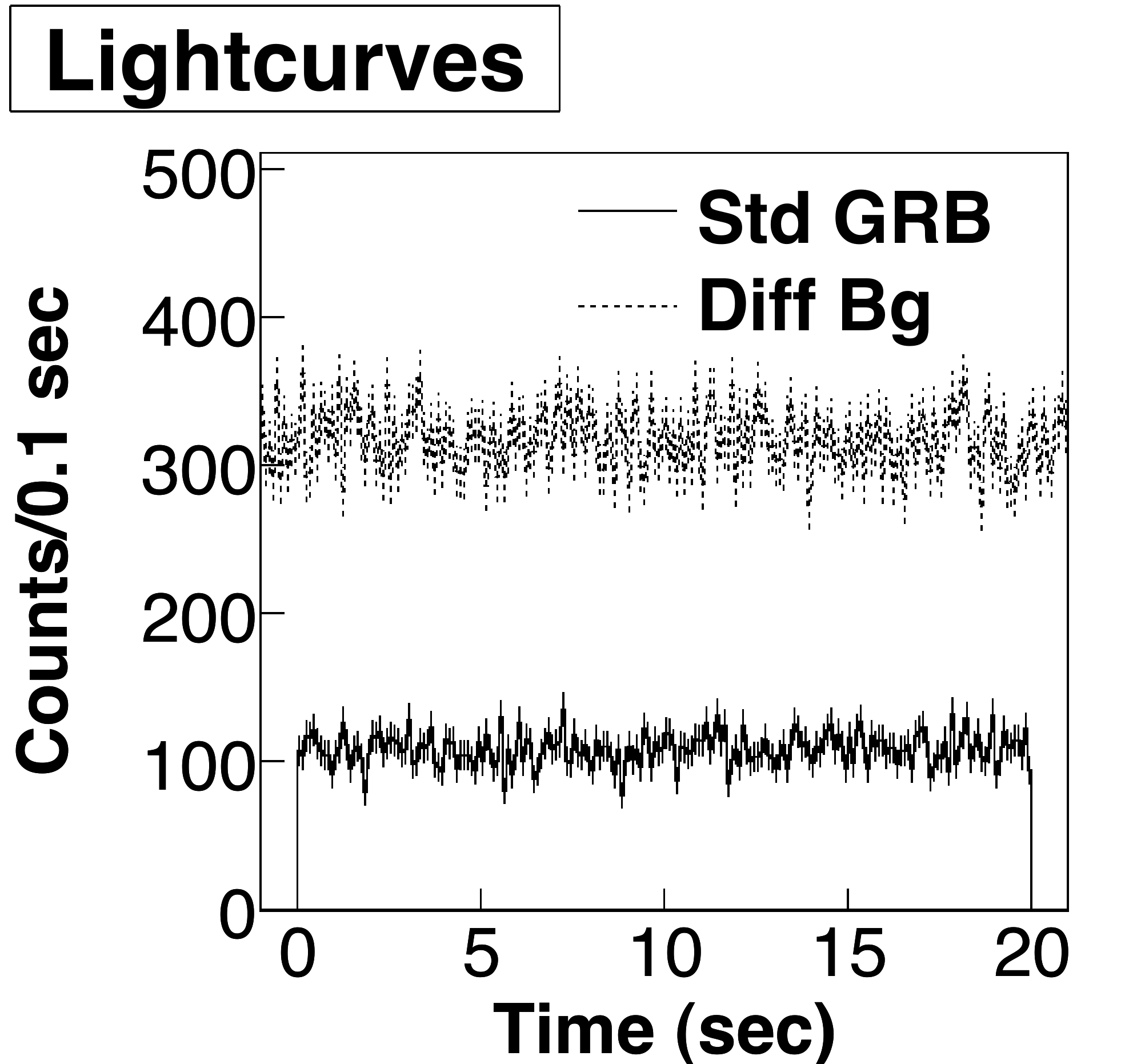} \includegraphics[width=4cm]{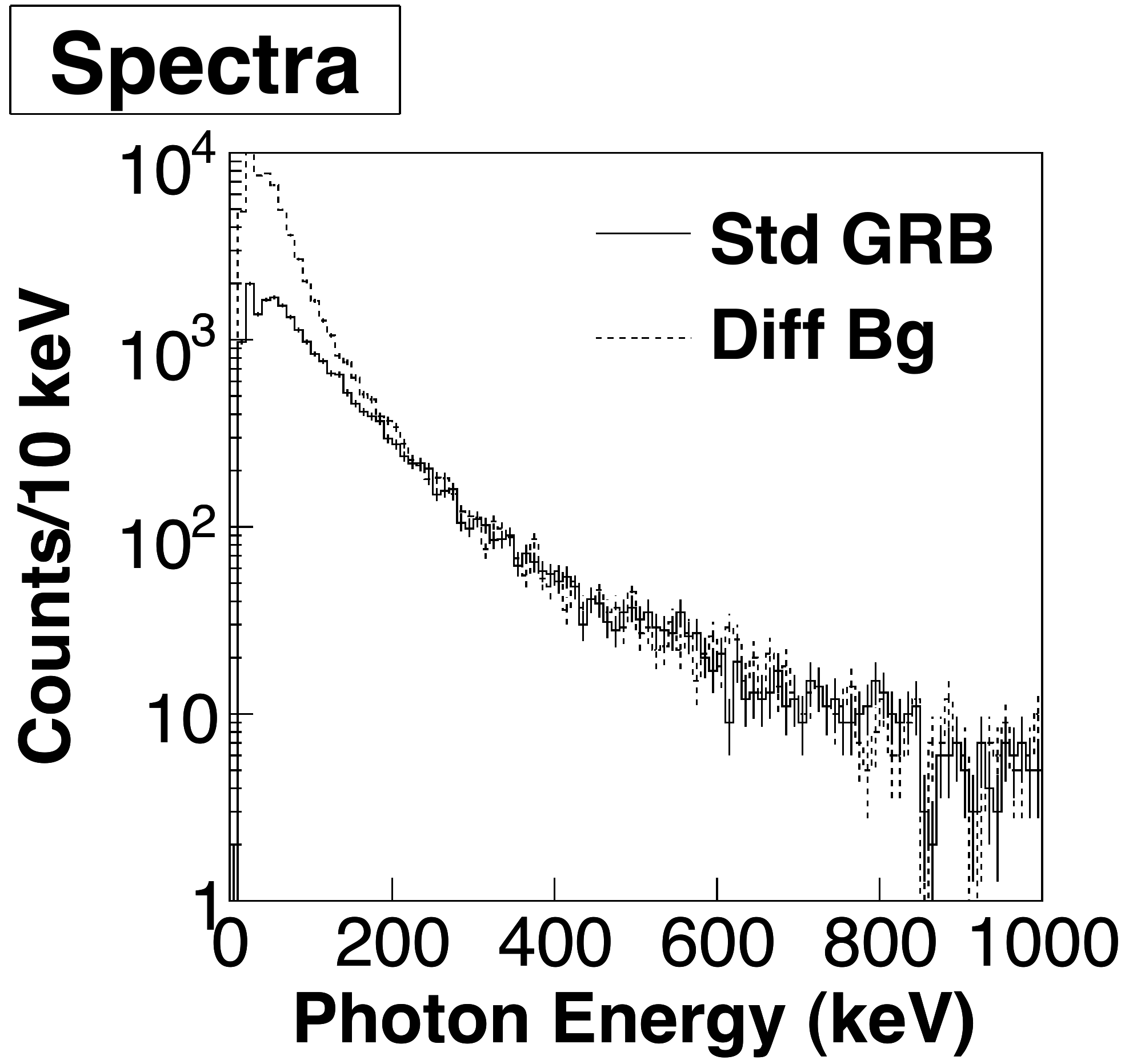}} 
\caption{Comparison between the flux of a long (20 sec) \textit{standard} strong GRB and the diffuse photon background. {\it Left}: Assumed lightcurve of the GRB signal ({\it solid}) and the diffuse photon background ({\it dotted}). All photons producing a hit above 5 keV in POLAR target have been included. {\it Right}: Spectrum for the same two sources.}
\label{fig:9}
\end{figure}

To study the background influence it is important to consider the duration of the burst. Unlike GRBs, where for a constant fluence the flux diminishes when the duration increases, the diffuse background produces a flux approximately constant with time. Since we accumulate the output of the scaler during the whole GRB, the shape of its lightcurve has no influence on the localization method. For simplicity we considered both the lightcurves of the burst and the background as flat. Figure \ref{fig:8} shows the lightcurve and spectra of the diffuse photon background together with a short GRB of 1 second duration and \textit{standard} Band parameters located at POLAR zenith. The equivalent plots for a 20 seconds long GRB are presented in figure~\ref{fig:9}.

We have simulated the diffuse $\gamma$-ray background using equation~\ref{eq:8} with a 10\% higher normalization factor as suggested in \cite{churazov}, and we calculated its SO. The result was added to the SOs of a series of \textit{standard} strong GRBs as the ones presented in figure~\ref{fig:4}. Then, a second diffuse $\gamma$-ray background was simulated and its SO was subtracted from the previous data successively with an excess of 8\%, 10\%, 20\%, and 30\% over its average value. The whole procedure was performed twice, using a 1 second and 20 seconds duration GRB, to reproduce a short and long GRB, respectively.

We observed that, while for a short GRB the background did not influence the performance of the localization method, its contribution was very important in the case of the long GRB, inducing an error of $\sim$50\% in $\mu_{100}$ if not subtracted. If the background is overestimated by a 8\% of its real level, the error transmitted to $\mu_{100}$ was $\sigma_{\mu_{100}}\le$10\%, increasing to $\sigma_{\mu_{100}}\le$14\% for a 10\% overestimation, and to $\sigma_{\mu_{100}}\sim$20\% and $\sigma_{\mu_{100}}\sim$30\% for a 20\% and 30\% overestimation, respectively. 

As a conclusion, background subtraction is important not only for the polarimetric performance of POLAR, but also for its capability of localizing GRBs. In the case of short GRBs the diffuse $\gamma$-ray background can be neglected, but for 20 seconds long GRBs the background calculation must not differ from the real background level by more than 8\%.  When the background is not strongly variable, it is sufficient to average two background measurements one taken immediately before and the other immediately after the GRB. 

\subsection{Variations of the GRB spectral shape}
\label{sec:influencespec}
The GRBs used for the NSO database have been simulated taking the Band parameters corresponding to a \textit{standard} strong GRB, summarized in the first row of table~\ref{tab:batse}. The NSOs and SOs depend on the spectrum since the penetration length of photons in the POLAR target is an exponential function of their energy. To study the possible influence of different GRB spectra on the localization method, several series of simulations have been performed. We selected from the BATSE catalog \cite{kaneko} the bursts with energy fluence $\rm{F_{tot}} \approx10^{-5}$ erg cm$^{-2}$ which had the Band parameters most different from our standard ones. The selection of GRBs studied is presented in table~\ref{tab:batse}.

\begin{table}[h]			       	 
\begin{center}
\begin{tabular}{lcccc}
\hline 
{\bf GRB} & ${\rm \bf{F_{tot}}}$ ($\times$10$^{-5}$)	& $\alpha$ & $\beta$ & {\bf E$_{\rm \bf peak}$} \\ 
	& [ erg cm$^{-2}$]	& 	& 		& [keV] 	\\ \hline
Standard	& 1.0 	 	& -1.0 	& -2.5	& 200 	\\
911104 	& 1.1 	& -0.6 	& -1.95	& 265 	\\
920718 	& 1.1 	& -1.09 	&  -3.06	& 192 	\\
930922	& 2.0 	& -1.40 	& -2.88	& 94 	\\
951016 	& 1.4 	& -1.61 	&  -2.02	& 118 	\\
980828	& 1.7 	& -0.25 	& -2.08	& 223 	\\
981130	& 1.1 	& -0.54 	& -2.25	& 649 	\\
\hline
\end{tabular}
\caption{Details of the spectrum of the \textit{standard} strong GRB compared to the bursts from BATSE catalog selected for studying the influence of the spectral shape into the localization procedure. From the BATSE catalog we selected the six bursts that,  having a similar total energy fluence, presented the most different Band parameters with respect to our \textit{standard} GRB.} 
\label{tab:batse}
\end{center}
\end{table}

We made a series of simulations locating each BATSE burst at several positions in the sky and applied to them the localization method to study the spread and bias of the reconstructed values, in the same way as presented in section~\ref{sec:verification}. When using a very low 5 keV energy threshold to the scalers, it was observed that the modification of the GRB spectral parameters introduced in some cases a large bias on the reconstructed $x$ and $y$ coordinates. The sign of the bias changed from GRB  to GRB, and its absolute value was the largest  (taking values up to 0.25) when $\alpha$ and $\rm{E_{peak}}$ strongly differed from the values used for the database, while variations on $\beta$ did not affect it significantly. The main reason is that most of the photons that POLAR detects, and which mostly define the shape of the SO, have low energies. Low energy photons do not penetrate far inside POLAR target, producing hits only on the side where the GRB is coming from. Although this makes them in principle very good to localize a GRB, the fact that POLAR cannot measure the spectral shape precisely enough makes it impossible to correct for the bias in the reconstructed coordinates for GRB spectrally very different from our standard one. 

\begin{figure*}[!ht]         	 
\centerline{\includegraphics[width=14cm]{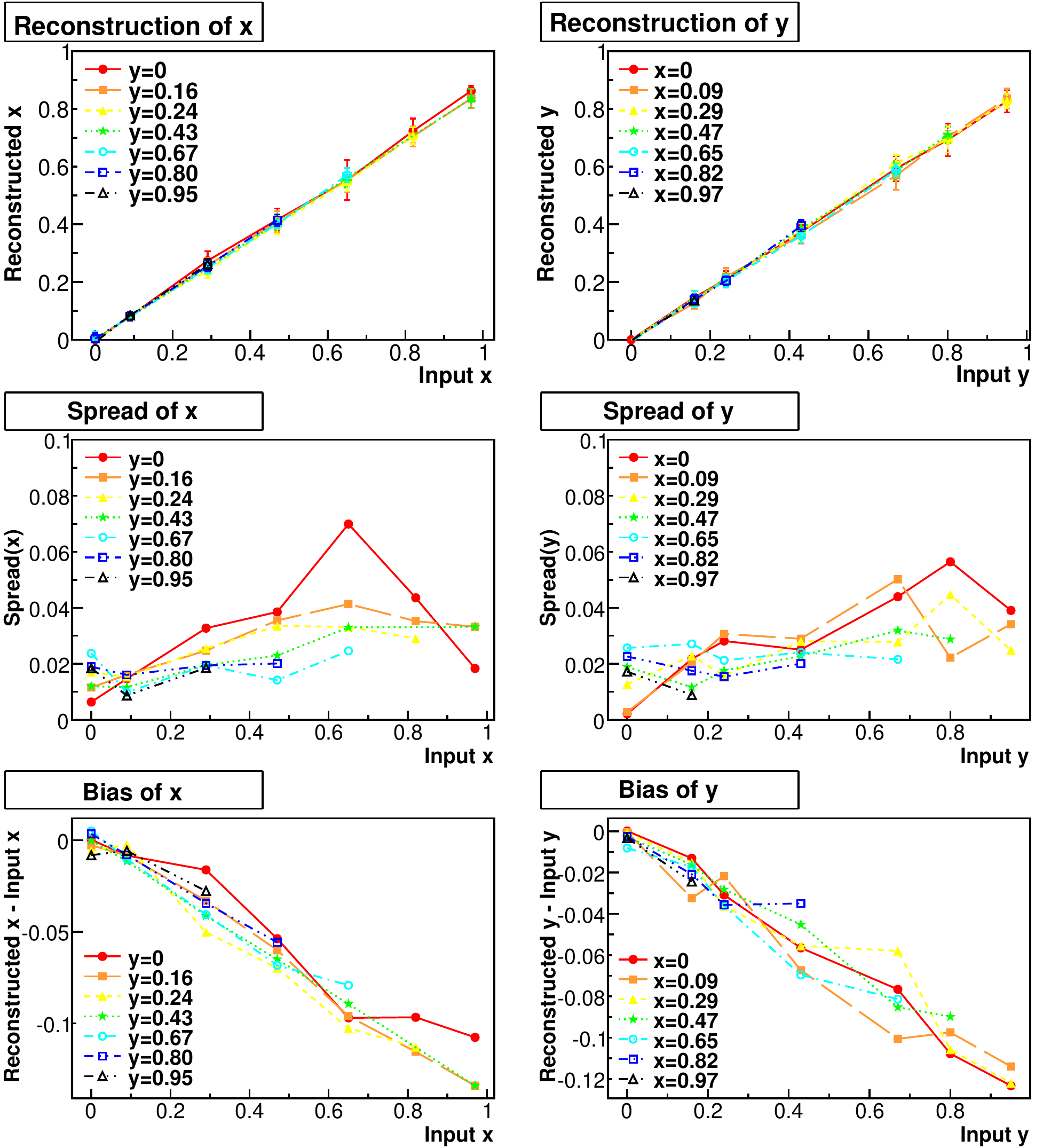}} 
\caption{Results from GRB 981130, simulated non-polarized and without background. \textit{Top left}: Linearity plot in $x$,  i.e. reconstructed $x$ vs. input $x$ for several fixed $y$ values.  Each point on this graph represents the average of the 10 simulations made at each position. \textit{Middle left}: Spread of the reconstructed $x$, i.e. the standard deviation of the 10 reconstructed $x$ values. \textit{Bottom left}: Bias in the reconstruction, i.e. the difference between the reconstructed average $x$ value and the input $x$. The plots on the right are the equivalent to the left ones, but calculated for $y$ at several fixed $x$ positions. In all the plots lines of different styles have been drawn to guide the eye.}
\label{fig:10}
\end{figure*}

One possible solution would be to measure the GRB spectrum with a small spectrometer that could be mounted somewhere close to POLAR. In such a case it would be enough to produce a new NSO database with the observed burst spectrum and apply the localization method to determine the GRB position with a few degrees uncertainty. When no spectrometer is available, the only alternative to mitigate the problem is, as we have done, to raise the energy threshold of the scaler, which in principle could be as low as 5 keV, to 50 keV. In this way we largely reduce the sensitivity of the method to spectral variations, so that the bias of the reconstructed coordinates is below 0.15 and the error inflicted into $\mu_{100}$  is below 10\% for all GRBs from table~\ref{tab:batse}. We show in figure~\ref{fig:10} the case of GRB 981130, for which the method performance was the worst of the six examples (the absolute values of the coordinates bias were the largest). 

Unfortunately, when reducing the method sensitivity to spectral variations, we cut out a large number of counts from the SOs and NSOs, diminishing the statistical power. As a consequence, the error produced on $\mu_{100}$ from the uncertainty in the localization of  GRBs with fluence {$ \rm F_{tot} = 10^{-6}$} erg cm$^{-2}$ and same spectral parameters as the database bursts, can reach 14\%. For weaker GRBs our technique will not be able to provide the position of the source with sufficient accuracy, especially taking into account the spectral variability.

\subsection{Satellite backscattering}
\label{sec:satellite}
Photons coming from the GRBs can reach POLAR indirectly after being backscattered in the spacecraft where the polarimeter is mounted. We have simulated a very simple description of the Chinese Tiang-Gong Space Lab as a "bottle-like" assembly of two cylinders (1.65 m radius, 4 m high,  and 1.4 m radius, 3.2 m high, respectively) joint by a 1.4 m long trunked cone. All volumes were simulated in aluminum with a thickness of 4 cm, so that the total mass of the spacecraft is $\approx$8 tons. POLAR was placed outside the Space Lab at 12.5 cm from its trunked cone part, pointing with the front of its target to the zenith. The GRB photons were uniformly illuminating an area of 2 m radius with POLAR at its center. A series of simulations with \textit{standard} strong GRBs at several positions in the sky, as done for figure~\ref{fig:4}, were produced and the localization method was applied. The results were found to be within errors the same as when the satellite was not considered. The 50 keV energy threshold applied in the scalers is rejecting most of the photons coming to POLAR from the satellite, since they have lost energy in the backscattering process.

\subsection{Statistical fluctuations in light collection}
\label{sec:lightcollection}
Plastic scintillators are a good choice to perform hard X-ray polarimetry because the probability that a photon of this energy range experiences Compton scattering in the target is very high. The price one pays in exchange is a low energy sensitivity, not enough to make spectroscopic studies. GEANT4 simulations of optical photon tracking show that about $\approx$60\% of the optical photons produced in a POLAR scintillator bar never reach the MAPM \cite{suarezgarcia}. The detector will be calibrated taking this into account but, due to the statistical fluctuations in the number of optical photons collected, the spectral resolution will be poor. We have simulated the scintillation process in a single bar and fired it with bunches of photons at various energies. For each run a gaussian fit of the photo-peak spectral line was performed. We found an expression to relate the $\sigma$ of this gaussian fit with the incoming energy of the photon, i.e., the broadening ($\sigma$) of the lines as a function of the deposited energy (${\rm E_{dep}}$, expressed in keV): 
\begin{equation}  		
\sigma = 0.215\cdot{\rm E_{dep} [keV]}+ 2.953 \mbox{ keV}.
\label{eq:9}
\end{equation} 

\noindent The main consequence of this line-broadening for POLAR measurements is that the signal collected at the MAPM anode does not correspond exactly to the deposited one. Therefore, the 50 keV threshold introduced at the scalers will be blurred.

The influence of the spectral line broadening into the GRB localization method has been studied. \textit{Standard} strong GRBs were simulated at different positions in the sky as it was done to produce figure~\ref{fig:4}. Before applying the localization method to these GRBs, each value of the energy deposition ${\rm E_{dep}}$ was substituted by a number that had been randomly generated following a Gaussian distribution with mean ${\rm E_{dep}}$ and sigma the result from equation~\ref{eq:9}. The application of the localization method to the modified data gave the same results as when ${\rm E_{dep}}$ was perfectly determined, with no increment on the coordinates uncertainty. 

\subsection{Sensitivity variations between MAPM anodes}
\label{sec:MAPMuniformity}
Two photons of the same energy can produce different signals if they come to two MAPM channels that do not have the same sensitivity. According to the H8500 MAPM data-sheet \cite{h8500} sensitivity differences up to a factor of 3 can exist between the anodes of the same MAPM. We have observed that the sensitivity usually varies following a smooth function with a maximum close to one of the corners of the MAPM, and monotonously decreases when moving away from that point. We have assigned to each of the 64 elements of a MAPM a number in the range from 0.4 to 1, distributed according to an example data-sheet, but with variations of the order of 5\% at each channel. Such a sensitivity mask was produced for each of the 25 MAPMs of POLAR, taking random orientations so that the corner with maximum sensitivity was not always at the same side of the target. In this study we have assumed that the average sensitivity is the same for all 25 MAPMs. The $40\times40$ mask created in this way was applied as a multiplicative factor to the energy depositions of a series of \textit{standard} strong GRBs like the ones from figure~\ref{fig:4}. No change in the localization uncertainty with respect to the results shown in figure~\ref{fig:4} was found. 

\section{Summary and conclusions}
\label{sec:summary}
Relevant information on the magnetic composition, geometric structures, and emission mechanism of GRBs, not accessible by spectroscopic and photometric approaches, can be obtained through polarization measurements of the GRB prompt emission. POLAR aims to determine the level of linear polarization of GRBs in the energy range from 50 to 500 keV by measuring the azimuthal distribution of the photons that Compton-scatter in its plastic scintillator target. To keep the design as simple and compact as possible, POLAR is totally devoted to polarimetry and is not optimized for imaging and spectroscopic purposes. The level of polarization of a GRB can only be determined when knowing its position in the sky.

We have presented a method, using only POLAR, able to estimate the position of GRBs with enough precision so that the error transmitted to the 100\% modulation factor due to the location uncertainty is kept below 10\% for strong GRBs of total fluence $\rm{F_{tot} \ge 10^{-5}}$ erg cm$^{-2}$. A scaler is connected to the output of each MAPM to record the number of GRB photons that produce at least one hit with energy above 50 keV. The outputs of the 25 scalers, accumulated over the whole duration of the GRB, present a pattern that depends on the direction where the photons are coming from. 

We have performed several series of simulations to estimate the capabilities of the GRB localization method and the influence of its uncertainty on the measured level of polarization. When simulating strong non-polarized GRBs with a fluence $\rm{F_{tot}}=10^{-5}$ erg cm$^{-2}$ and a similar spectral shape to the ones used to create the database, the position coordinates could be determined with a maximal error of 0.07, equivalent to being able to determine the position in spherical coordinates ($\theta_\gamma$, $\phi_\gamma$) within $\sim 5^\circ$ for $\theta_\gamma > 30^\circ$. This would inflict an error in the $\mu_{100}$ of the measured GRB of less than 6\%. Considering GRBs of similar strength but different spectral shapes, the coordinates could be determined with a maximal error of 0.14, keeping the error on $\mu_{100}$ below 10\%. For weaker GRBs of fluence $\rm{F_{tot}}\approx10^{-6}$ erg cm$^{-2}$ the error in $\mu_{100}$ will be below about 20\%.

Several effects that could potentially modify the performance of the GRB localization technique have been studied. It was found that the influence of the diffuse photon background will be negligible for short GRBs, but it needs to be subtracted from long ones. Its estimation in the later case should be done with a precision better than 8\% around the real value. Neither the level of polarization of the source, nor the presence of a spacecraft behind POLAR that backscatters part of the GRB photons show any influence in the output of the localization method. The limited energy resolution due to light collection inefficiencies and MAPM non-uniformities will blur the 50 keV energy threshold but this will not diminish the localization capabilities of our method. 

The GRB localization technique will provide enough accuracy to allow for the measurement of the GRBs polarization level with only a small increase on its error. For a GRB of total fluence $\rm{F_{tot}}=10^{-5}$ erg cm$^{-2}$, the added error is negligible at low polarization levels, compared with the one from the measured modulation factor itself. At large polarization levels the error from the localization is larger than the one from the measured modulation factor, but all together reaching no more than 15\% of the measured polarization. 

The method presented here can be experimentally tested when uniformly illuminating the complete POLAR target with a radioactive source to be placed at various locations. Such a kind of test will be performed when the POLAR engineering qualification model will be ready. In flight, if other instrument can provide the position of one or several GRBs detected by POLAR, a direct comparison with the output of the localization method can be done. In general, the precision in the localization provided by instruments with imaging capabilities will be much better than that of POLAR alone, and preferred to it. The POLAR localization capability will be useful for those cases where no other instruments are simultaneously observing the same field of view.

\section{Acknowledgements}
We wish to thank the technical staff at the University of Geneva, in particular F. Masciocchi and F. Cadoux, for their crucial contribution to the POLAR mechanical design, and for producing figure~\ref{fig:1} for this paper. This work was funded by the Fonds National Suisse pour la Recherche Scientifique.



\end{document}